\documentclass[conference]{IEEEtran}
\IEEEoverridecommandlockouts
\usepackage{cite}
\usepackage{balance}
\usepackage{amsmath,amssymb,amsfonts}
\usepackage{multirow}
\usepackage{algorithmic}
\usepackage{graphicx}
\usepackage{textcomp}
\usepackage{xcolor}
\usepackage{url}
\graphicspath{{Figures/}}
\usepackage{subfig} 
\usepackage{fancyhdr}
\usepackage{amsmath}
\usepackage{graphicx}
\usepackage{fancyhdr}
\usepackage[top=0.7in, bottom=0.6in, left=0.7in, right= 0.7in]{geometry}
\usepackage{tikz}
\usepackage{lipsum}

\fancypagestyle{FirstPage}{
\lfoot{978-1-7281-5628-6/20/\$31.00 \copyright~2020 IEEE}
}

\def\BibTeX{{\rm B\kern-.05em{\sc i\kern-.025em b}\kern-.08em
    T\kern-.1667em\lower.7ex\hbox{E}\kern-.125emX}}
\begin{document}

\title{Detecting Abnormal Traffic in\\ Large-Scale Networks
}

\author{\IEEEauthorblockN{Mahmoud Said Elsayed}
\IEEEauthorblockA{\textit{School of Computer Science} \\
\textit{University College Dublin}\\
Dublin, Ireland \\
mahmoud.abdallah@ucdconnect.ie}
\and
\IEEEauthorblockN{Nhien-An Le-Khac}
\IEEEauthorblockA{\textit{School of Computer Science} \\
\textit{University College Dublin}\\
Dublin, Ireland \\
an.lekhac@ucd.ie}
\and
\IEEEauthorblockN{Soumyabrata Dev}
\IEEEauthorblockA{\textit{School of Computer Science} \\
\textit{University College Dublin}\\
Dublin, Ireland \\
soumyabrata.dev@ucd.ie}
\and
\IEEEauthorblockN{Anca Delia Jurcut}
\IEEEauthorblockA{\textit{School of Computer Science} \\
\textit{University College Dublin}\\
Dublin, Ireland \\
anca.jurcut@ucd.ie}

}

\maketitle

\thispagestyle{FirstPage}

\begin{abstract}
With the rapid technological advancements, organizations need to rapidly scale up their information technology (IT) infrastructure \emph{viz.} hardware, software, and services, at a low cost. However, the dynamic growth in the network services and applications creates security vulnerabilities and new risks that can be exploited by various attacks. For example, User to Root (U2R) and Remote to Local (R2L) attack categories can cause a significant damage and paralyze the entire network system. Such attacks are not easy to detect due to the high degree of similarity to normal traffic. 
While network anomaly detection systems are being widely used to classify and detect malicious traffic, there are many challenges to discover and identify the minority attacks in imbalanced datasets. In this paper, we provide a detailed and systematic analysis of the existing Machine Learning (ML)  approaches that can tackle most of these attacks. Furthermore, we propose a Deep Learning (DL) based framework using Long Short Term Memory (LSTM) autoencoder that can accurately detect malicious traffics in network traffic. We perform our experiments in a publicly available dataset of Intrusion Detection Systems (IDSs). We obtain a  significant improvement in attack detection, as compared to other benchmarking methods. Hence, our method provides great confidence in securing these networks from malicious traffic. 
\end{abstract}


\section{Introduction}
\label{sec:intro}


Intrusion is the main problem of security breaches, where a malicious user can damage or steal vital information of the network system in a short time.  Moreover, it can cause further financial losses and huge damages in IT critical infrastructure. For example, \$350M and \$70M are the size of loss caused by Yahoo and Bitcoin data breach, respectively~\cite{larson2016distributed}. The intruder techniques have been evolved using sophisticated tools to create attacks, exploiting vulnerabilities in server protocols~\cite{jurcut2017, jurcut2019}. 

Primarily, the four popular attack classes that can cause extensive damage to any network environment include:
\begin{itemize}
    \item Probe: It is not considered an attack by itself, but it is a primary step for any attacker to get more information about any network vulnerabilities that will allow for a future attack.
    \item R2L: The R2L attack involves the attacker to access the victim machine using some vulnerabilities from the probing phase to get illegal access or local right permissions.
    \item U2R: In this type of attack, the attacker attempts to get full privilege in the victim machine, who has only local access on it.
    \item Denial-of-Service (DoS): Finally, in this type, the attacker attempts to overwhelm or flood the target machine or network, with a view to disturb its normal operation. 
\end{itemize}

The traditional defense techniques, \emph{i.e.} firewall, fail to detect intrusion attacks inside the network system due to its nature and its specific features~\cite{dong2016detection}. Recently, several ML models~\cite{imamverdiyev2018deep, dawoud2018deep, mohammadi2017new, naseer2018enhanced, elsherif2018automatic} have been developed to provide a framework for detecting anomalies in network traffic data. These frameworks are evaluated and benchmarked against standard datasets that are publicly available. The KDDCUP'99~\cite{tavallaee2009detailed} and its modified version NSL-KDD dataset~\cite{dhanabal2015study} are the most widely used datasets in anomaly detection application domain. The NSL-KDD\footnote{The NSL-KDD  is available for download here: \url{https://www.unb.ca/cic/datasets/nsl.html}.} dataset is a modified version of KDDCUP'99 dataset, attempting to improve the inherent demerits of the original dataset.

The current state-of-the-art algorithms provide a rigorous framework for malicious traffic detection in the NSL-KDD dataset. However, these methods suffer from several drawbacks. Firstly, their performance is good only in the training set of the dataset, while the performance in the corresponding testing set is poor.  Secondly, the detection accuracy of these ML-based methods for certain types of attack \textit{viz.} U2R, and R2L is lower than other traffic types. This is returned to the unbalanced nature of the dataset -- the number of U2R and R2L observations in the dataset are much lower than other traffic types. 
We identify these gaps in the literature and address them in this paper by proposing a Deep Neural Network (DNN) for efficient detecting network attacks. The proposed model has been developed by integrating the LSTM layers with autoencoder to model the normal trafﬁc data. 

The main contributions of this paper are as follows -- (a) we review the related methods based on ML and discuss their limitations; (b) we propose a DL based model for encoding the input feature space with a high discriminatory nature and propose an approach based on LSTM to detect malicious traffic in the network; and (c) we benchmark several state-of-the-art ML models for detecting attacks in the publicly available NSL -KDD dataset. Our proposed method for detecting network attacks has the best performance in the existing benchmark dataset. 

The structure of the remaining part of the paper is as follows: Section~\ref{sec:rworks} provides a systematic review of the various learning models for detecting attacks in large-scale networks. We propose our DL model for encoding the input feature space and the binary classification framework in Section~\ref{sec:proposed}. Section~\ref{sec:results} presents our benchmarking results in the publicly available NSL-KDD dataset. Finally, Section~\ref{sec:conc} concludes the paper and discusses our future work. 

\section{Detecting attacks in large-scale networks}
\label{sec:rworks}

In this section, we provide a detailed discussion of the various ML and DL models that have been proposed for detecting sophisticated attacks in large-scale network environments. With the advent of DL based models, the accuracy in detecting attacks has further improved. The DL based methods are useful, because the discriminative features are automatically generated, without the use of generating hand-crafted features.

\subsection{Related Work}


In this section, we discuss some related works on anomaly detection systems on the NSL-KDD dataset. 

Dhaliwal in~\cite{dhaliwal2018effective} introduced an intrusion detection module based on XGBoost algorithm to detect malicious activities within the network environment. The calculated accuracy of the classifier model is 98.40\% and F1-Score of 98.76\%. However, the authors combined both training and testing files of the NSL-KDD dataset to create heterogeneous data that contains the same attack distribution. Ever \textit{et al.} in~\cite{ ever2019classification} presented intrusion classification model using NSL-KDD dataset. The authors employed three different ML algorithms; Decision Tree (DT), Support Vector Machine (SVM), and back-propagation neural network. Two different experiments are repeated for each deployed technique. In the first experiment, 60\% of the dataset is used for training, and the rest of 40\% is used for testing. While in the second experiment, 70\% of instances are used for the training set and 30\% for testing. The results show that the high accuracy is achieved from DT model with percentages of 99.84\% and 99.81\% for each experiment, respectively. However, the authors evaluated their models using the NSL-KDD training file only without looking at the separated testing portion of the dataset.

Alshamrani \textit{et al.} in~\cite{alshamrani2017defense}, proposed a defense system to mitigate DDoS attack inside the Software Defined  Network (SDN). The represented module is evaluated using three different ML algorithms Naïve Bayes (NB), J48, and SVM on the NSL-KDD dataset. The authors satisfied high efficiency using SVM with an approximate accuracy of 99\%. Using a separated portion of the testing data for the model evaluation gives poor accuracy results as the new attack records in the testing file are different from those used for the training. Ingre \textit{et al.}~\cite{ingre2017decision} implemented an intrusion detection system based on DT classifier against cyberattacks using NSL-KDD. They applied a random sampling technique to address the problem of minor classes of R2L and U2R in the NSL-KDD dataset. However, the overall accuracy for the five-class classification is 83.7\%, where the detection rate for R2l and U2R is 39.9\% and 8.5\%, respectively. 

Latah \textit{et al.}~\cite{latah2018efficient} proposed a five-stage hybrid classifier system to enhance the detection rate against malicious traffics inside the network. The model combines three different ML classifiers, including the K-Nearest Neighbor approach (KNN), Extreme Learning Machine (ELM), and Hierarchical Extreme Learning Machine (H-ELM). The authors used five classification layers where only one specific attack can be detected at each layer. Therefore the experiment is executed five times to detect normal and all attack categories. The overall accuracy of the presented approach is 84.29\%, while the percentage of precision, recall, and F1-score is 94.18, 77.18, and 84.83, respectively. Prasath \textit{et al.}~\cite{prasath2019meta} proposed a Novel Agent Program (NAP) framework to secure a communication model of the virtual switches in the network. The meta-heuristic Bayesian network classification (MHBNC) approach is used to classify the incoming packets into normal or attack traffic. The proposed MHBNC model has achieved an overall accuracy of 82.99\%. Besides, the realized precision, recall, and f-score is 77\%, 74\%, and 75\%, respectively.   

Wang \textit{et al.}~\cite{wang2018lesla} proposed a detection scheme to detect the control saturation attack in the e-health monitoring system. The authors implemented a contrastive pessimistic likelihood estimation (CPLE) mechanism in semi-supervised learning using the NSL-KDD dataset. The model satisfied high efficiency when 500 labeled, and 120000 unlabeled samples are used for the benchmark dataset. The overall accuracy of the presented method is 80.93\%. In~\cite{tang2016deep}, the authors presented a flow-based detection approach using Self Organizing Maps (SOMs) classifier in the network environment. The evaluated accuracy of the proposed model in the testing phase is 74.67\%. For overall evaluation, the precision, recall, and f-measure obtained from the conducted model are 83\%, 76\%, and 75\%, respectively.

\subsection{Limitations in the current ML-based methods}
\begin{figure}[htb]
  \begin{center}
    \includegraphics[width=3in]{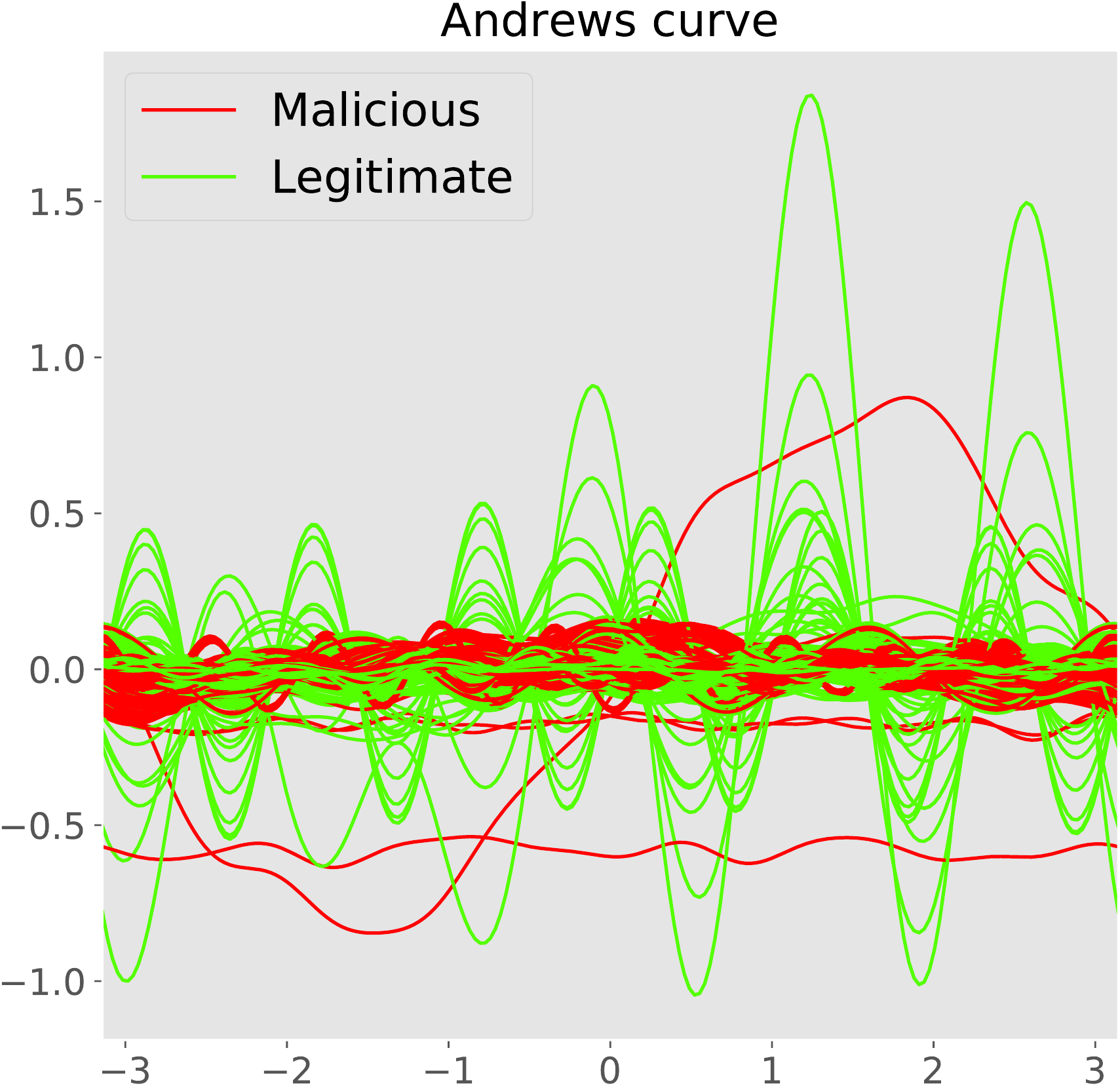}
  \end{center}
  \caption{We demonstrate the Andrews curve for the NSL-KDD dataset. We plot the legitimate and malicious observations in green and red curves respectively.}
  \label{fig:and-curve}
\end{figure}

ML techniques have been widely deployed in intrusion detection systems to recognize attack threats~\cite{sultana2019survey, kuang2014novel,garg2017novel,li2014new}. However, most of these works attempt to learn the feature representation in network traffic data for effective classification. However, it is not easy to manually handcraft and extracts the discriminatory features in intrusion detection systems. Firstly, the nature of the attacks evolve every day, and the attacker's techniques change with time. Secondly, the features which are extracted for one category of attack, may not necessarily be suitable for other attack classes. As a result, selecting the significant features to identify the attack from network traffic is a cumbersome task. Therefore, existing attack detection techniques fail to discover all types of attacks. In addition, there is a high degree of non-linearity in the dataset, and therefore the traditional ML-based methods fail to classify the normal and malicious data types~\cite{elsayed2019machine}. Elsayed \textit{et al.} further established this fact in~\cite{elsayed2019machine} by generating the Andrews curve for the NSL-KDD dataset. The Andrews curve represents a high-dimensional feature space in the form of a finite Fourier series. This provides a visual understanding of the internal structure of the dataset. Figure~\ref{fig:and-curve} shows the Andrews curve for the NSL-KDD dataset. Each curve in Fig.~\ref{fig:and-curve} represents an observation in the dataset. We observe that the two labels are not clearly grouped in two separate streams. The legitimate and malicious data curves are tangled with each other, indicating a high degree of inherent non-linearity in the feature space. Therefore, traditional ML techniques fail to capture the non-linearity in such datasets. 

, 

In this paper, we therefore consider DL techniques using LSTM-autoencoder to model the normal traffic data. This helps us in proposing a robust framework for detecting attacks in network traffic. We formulate our problem as a binary classification problem, wherein we classify any data samples into \emph{normal} and \emph{malicious} type. Unlike other ML approaches, our proposed model can automatically learn the discriminatory features from the network traffic data. 



\section{Proposed model}
\label{sec:proposed}
This section presents the framework elements and the system architecture to build a better IDS model.

\subsection{Modeling the normal traffic data}
In this section, we introduce our proposed architecture for detecting network attacks. 
We know that DL techniques use multiple processing layers to model the input feature accurately. This is advantageous as compared to traditional hand-crafted feature descriptors. Where DL techniques can automatically extract the discriminatory features and provide much better performance compared to classical ML models. Our proposed approach can estimate a good representation of the input feature space.


\begin{figure}[htb]
  \begin{center}
    \includegraphics[width=2.3in]{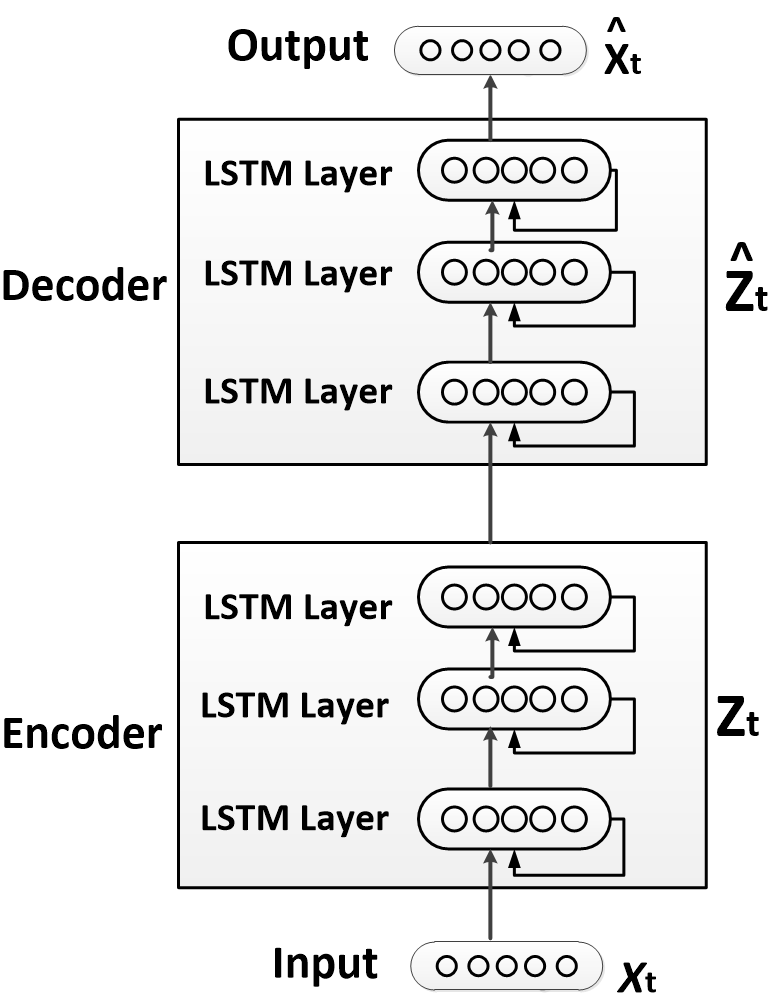}
  \end{center}
  \caption{Proposed model to encode the input features. We use blocks of encoder and decoder comprising LSTM layers.}
  \label{fig:architecture}
\end{figure}

Figure~\ref{fig:architecture} describes our proposed architecture to model the normal network data. 
We use the Recurrent Neural Network (RNN) and autoencoder architectures based on the nature of the input data, where the temporal correlations of network traffic often generate time-series data~\cite{tang2018deep}. For this reason, we used the RNN-based approach to solve the problem of simple feed-forward neural networks, since RNN considers the previous output and the current input at each stage. In addition, RNN has been applied efficiently in many of anomaly detection systems~\cite{elsayed2020ddosnet1}. Training the model with such methods can minimize the loss and further, it can provide high performance. Additionally, the autoencoder has the advantage in number of classification problem. The reason that we decided to use the autoencoder in  our proposed model for anomaly detection is the fact that the autoencoder is trying to learn the best parameters to reconstruct the input at the output layer. Moreover, we adapted the LSTM algorithm for our model to solve the issues of the standard RNN technique, such as vanishing and exploding gradient problems~\cite{kim2016long}.


Our model uses LSTM-autoencoder to learn the representations of the network dataset in an unsupervised fashion. Our model contains multiple layers of encoder and decoder stages, and each stage consists of multiple LSTM units. The input data $\mathbf{X}_t$ is encoded via the encoder block to generate the fixed range feature vector $\mathbf{Z}_t$. The input data $\mathbf{X}_t \in {\rm I\!R}^{122 \times 1}$ is the initial encoded feature vector generated from the dataset. We set \texttt{timestamp} = $1$ for our LSTM blocks. We used the LSTM blocks for individual events, and not for time series. The encoder block sequentially reduces the dimension of the $122$ dimension initial feature vector. The dimensions are reduced to $32$, $16$, and $8$, after the first, second, and third layers of the encoder, respectively. The final encoded feature vector $\mathbf{Z}_t \in {\rm I\!R}^{8 \times 1}$ represents the compressed input data.

The encoded data is then fed into the decoder block for generating the output feature vector. We represent the input feature vector of the decoder block as $\widehat{\mathbf{Z}_t}$. The layers in the decoder block are arranged in the reverse order as that of the encoder layers. The encoded features $\widehat{\mathbf{Z}_t}$ are then fed via a series of LSTM blocks to generate the output feature vector $\widehat{\mathbf{X}_t}$. The dimensions are increased to $8$, $16$, and $32$, after the first, second, and third layers of the decoder, respectively. Finally, the final layer of the decoder block is fed to a fully connected layer to generate the output feature vector $\widehat{\mathbf{X}_t}$. We attempt to reconstruct this output feature vector $\widehat{\mathbf{X}_t}$ to be similar to the input feature vector $\mathbf{X}_t$. We use the Mean Square Error (MSE) to calculate the estimation error between input data $\mathbf{X}_t$ and output representation $\widehat{\mathbf{X}_t}$. 

The reconstruction error for normal traffic data will be less, as compared to that of anomalous traffic data. This behavior will greatly help in detecting the anomalous traffic, as its corresponding error value will be considerably higher.

\subsection{Anomaly detection framework}
We train our DL model using normal data only without consider any anomalies traffic data. We compute the $\ell_2$-norm error between the original feature $\mathbf{X}_t$  and the output feature $\widehat{\mathbf{X}_t}$ to compute the reconstruction error. The $\ell_2$-norm error $e = \| \mathbf{X}_t - \widehat{\mathbf{X}_t}\|^2$ will be low for \emph{normal} traffic data, and high for \emph{anomalous} traffic data. Therefore, we use the reconstruction error to obtain the realistic threshold for the binary classification of \emph{normal} and \emph{anomalous} traffic data. The threshold value is used as a decision boundary for detecting anomalous data. The observations that have a reconstruction error greater than the threshold value will be classified as anomalous, while the samples that have  reconstruction error less than the threshold are classified as normal traffic data.


\section{Evaluation and Results}
\label{sec:results}
\subsection{Dataset}
Selecting the proper datasets is very important for evaluating intrusion detection performance. There are different methodologies to generate the intrusion detection datasets~\cite{ahmad2018performance} such as i) test-bed dataset, ii) sanitized dataset, iii) standard dataset. However, it is difficult to generate real network traffic from the first two methods. 
In addition, various attack categories are needed to create a real dataset that can reflect the real network traffic. To address the previous limitation, we used NSL–KDD dataset to evaluate the proposed DL model. The NSL-KDD dataset is a new modified version of the original datasets KDDCUP'99. It was generated typically to remove the inherent problems in the original dataset. The NSL-KDD dataset is now the \emph{de facto} benchmark dataset for the purpose of analysis network intrusion.


The NSL-KDD dataset is divided into training and testing sets for the purpose of ML-Algorithms evaluation. There are two variants of training and testing sets. The two files KDDTrain+ and KDDTest+ indicate the full version of the dataset with all observations. Smaller subsets of the dataset are available and are referred to as KDDTrain-- and KDDTest-- datasets. The NSL-KDD dataset contains $22$ specific types of attacks with $17$ additional types in the testing set that do not appear in the training set. Each observation in the dataset belongs to one of the five main categories: Normal, DoS attack, probing attacks, U2R, and R2L. In this paper, we focus our attention on a binary classification problem, wherein we classify each observation as \emph{normal} or \emph{anomalous} traffic data. The observations belonging to DoS, probing, U2R, and R2L are categorized as anomalous traffic data.


\subsection{Loss trend of our proposed model}

All the experiments were implemented in Python programming language using various libraries such as Scikit-Learn, Keras, and Tensorflow. We also execute all the experiments using a workstation machine that has the following properties: Intel(R) UHD Graphics 620, I7-8650UCPU @ 1.90GHz (8 cores), 2.1GHz, Windows 10 pro 64-bit with 16 GB of RAM.

We model the training file of NSL-KDD dataset as a fixed-length feature vector. All input features which are numeric in nature are normalized to the range $[0,1]$. We transform the categorical features into numerical values using one-hot encoding. The final input feature $\mathbf{X}_t$ is a $122$-dimension feature vector. We use the generated feature $\mathbf{X}_t$ to train our model, in this case the reconstruction loss is very low. We conducted different experiments to get the best values of hyper-parameters for model initiation. During these experiments, we change the value of learning rate, hidden layers size, epochs, and batch size and select the values that provide a high accuracy rate. We use a learning rate of $0.0001$, batch size of $32$, and train the model using Adam optimizer. We use Rectified Linear Unit (ReLU) as the activation function to overcome the vanishing gradient problem. 

\begin{figure}[htb]
  \begin{center}
    \includegraphics[width=2.5in]{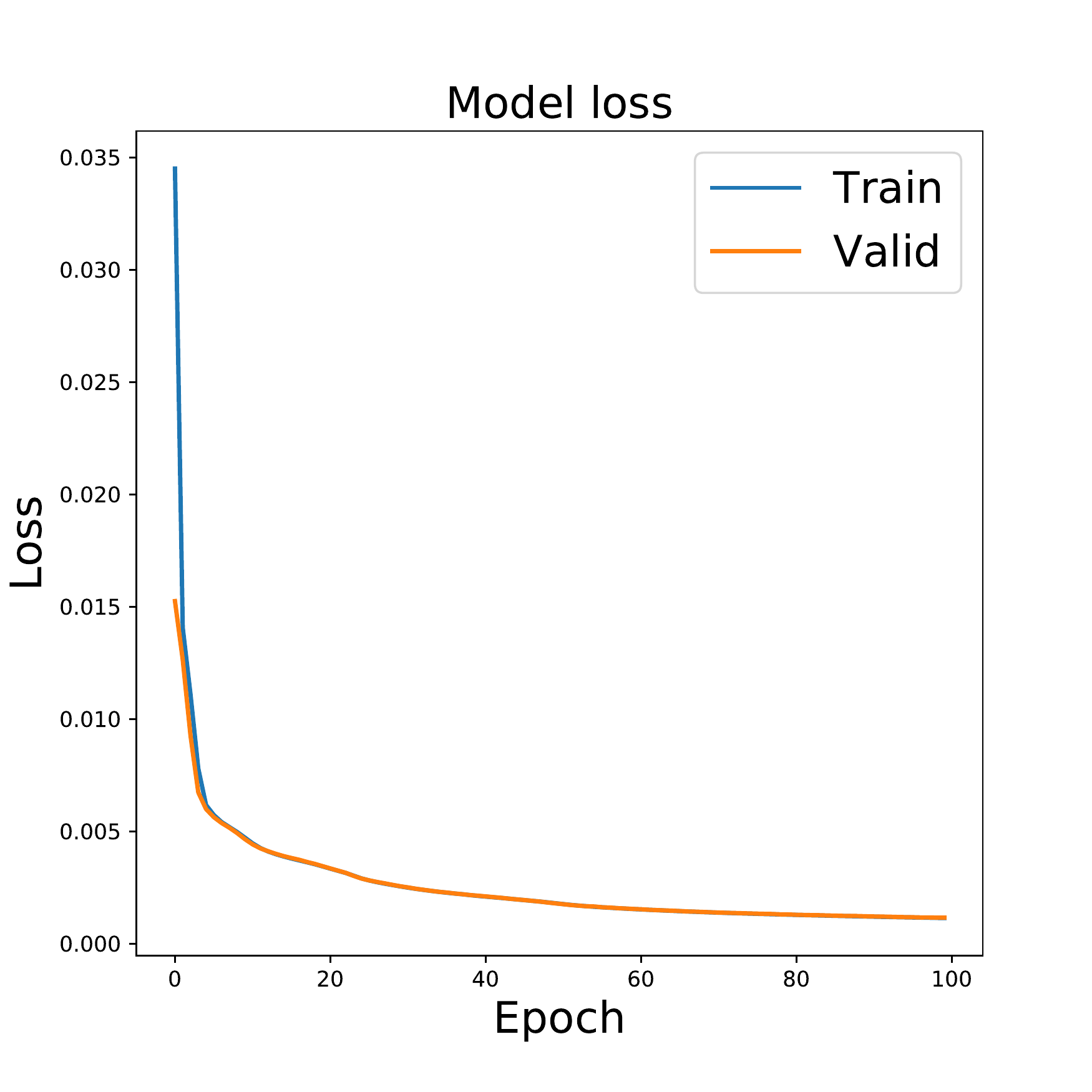}
  \end{center}
  \caption{Trend of training and validation loss over the number of epochs.}
  \label{fig:loss}
\end{figure}

Table~\ref{table:Parameters} summaries the choice of different hyper-parameters. The trend of training and validation loss over the number of epochs is depicted in Fig.~\ref{fig:loss}. We observe that the loss trend is similar for training and validation sets, and converges after a few hundreds of epochs.


\begin{table}[htb]
\centering
\small  
\begin{tabular}{l|c}
\textbf{ Parameters} & \textbf{ Best Values} \\
\hline 
Hidden layers & 3 \\
Hidden  layer  size  (neurons)& 32, 16, and 8 \\
Optimizer& Adam \\
Loss  function  & MSE \\
Activation  function  & ReLU\\
Batch  size & 32\\
Learning rate & 0.0001\\
Number of epochs& 100\\

\hline 
\end{tabular}
\caption{We mention the values of the several hyper-parameters.}
\label{table:Parameters}
\end{table}

\subsection{Detecting different attack types}

\begin{figure}[htb]
  \begin{center}
    \includegraphics[width=3in]{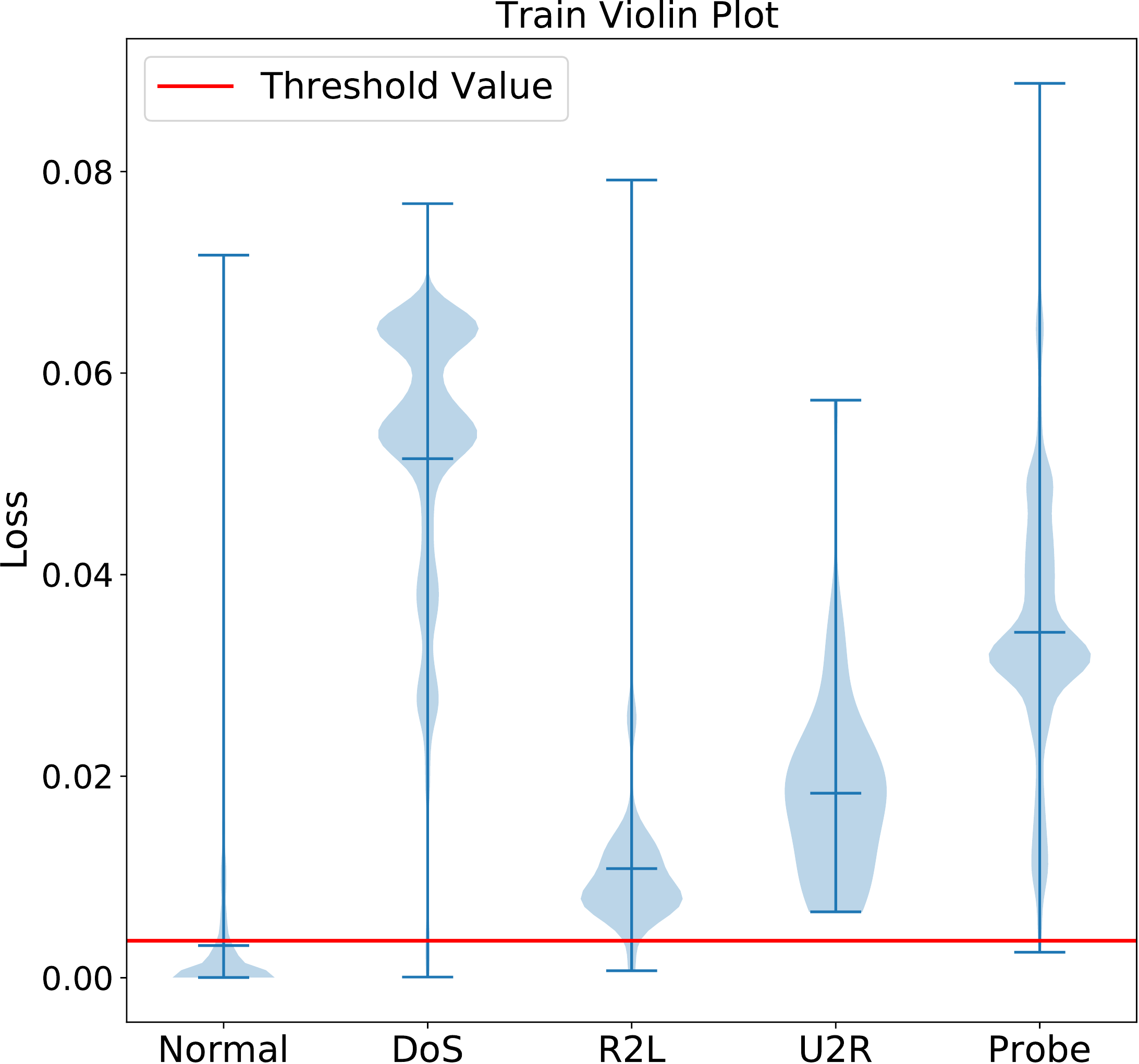}
  \end{center}
  \caption{Distribution of the reconstruction errors for normal traffic and various attack types traffic data. This violin plot represents the probability density of the data smoothed by a kernel.}
  \label{fig:dist-error}
\end{figure}

Our trained approach can model the normal traffic data with the least reconstruction error. This serves as the measure to identify the anomalous traffic data. We compute the distribution of the reconstruction loss for all the observations in the training set. We pick different threshold values and select the optimal value, which give best results for the detection rate. Figure~\ref{fig:dist-error} describes the reconstruction loss based on the five different labels. 

We observe that the errors of the four types of anomalous traffic data is much higher than that of the normal traffic data. We mark this threshold in Figure~\ref{fig:dist-error} with a red line. We set a threshold of $0.00368$ to differentiate the normal and anomalous traffic data. In the testing set, we use this threshold to differentiate the normal and anomalous for each data sample. The observations that have a reconstruction error greater than the threshold are categorized as anomalous traffic; whereas the observations less than the threshold are categorized as normal traffic.

\begin{figure}[htb]
  \begin{center}
    \includegraphics[width=0.45\textwidth]{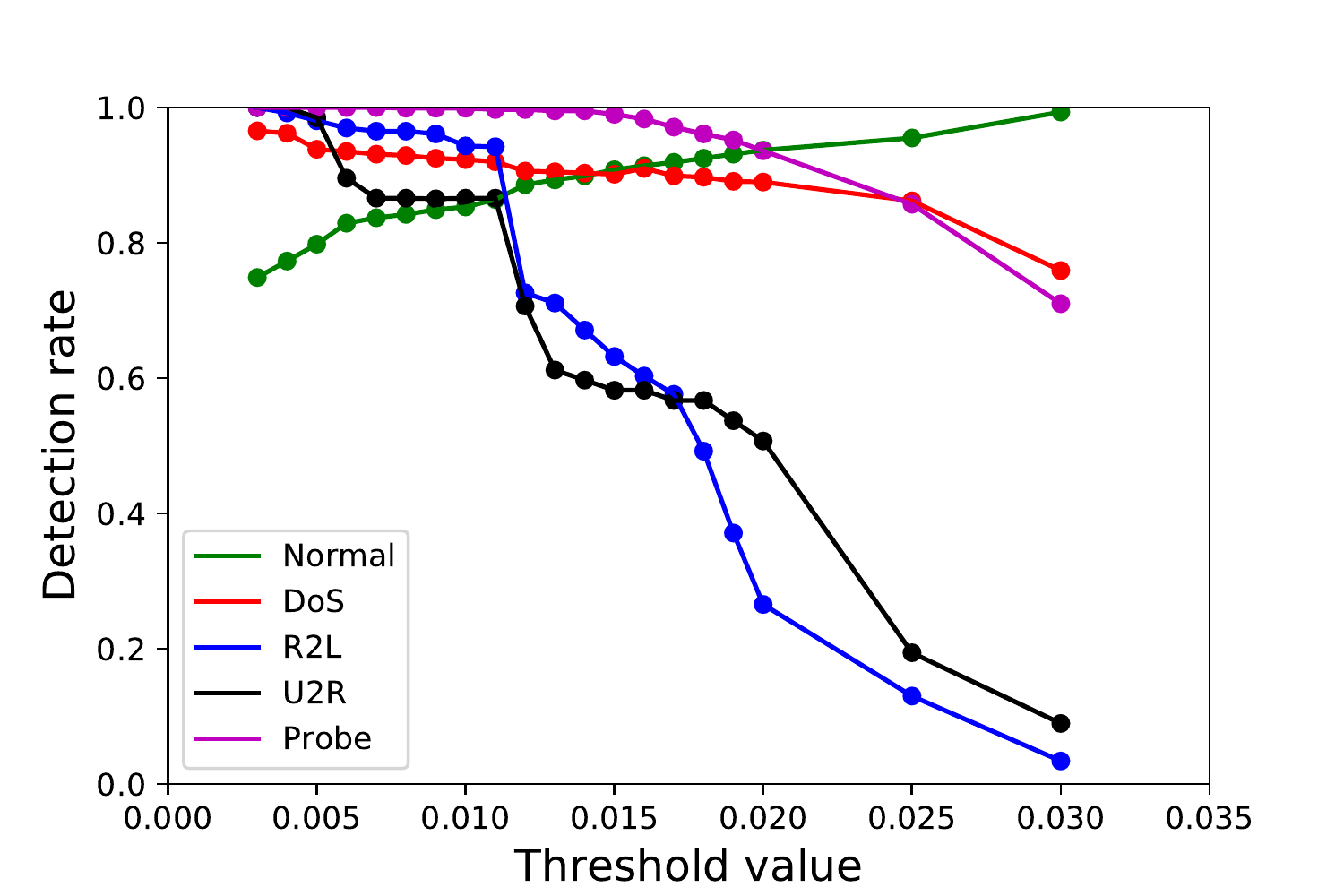}
  \end{center}
  \caption{Detection rate of the five different traffic Types with different threshold values.}
  \label{fig:threshold}
\end{figure}

Figure~\ref{fig:threshold} describes the different detection rate of the five data classes with different threshold values. We notice that the change in the threshold value does not cause a significant change in DoS and Probe attacks. This is because both DoS and Probe categories are commonly more different from normal traffic patterns~\cite{gogoi2013mlh}. In contrast, the last two attack classes (R2L and U2R) have high similarity to the normal connections, and therefore, at a specific threshold, we found unexpected decline in U2R and R2L instances. Since the minority attacks can cause a serious risk on the system, it is essential to identify them early before normal traffic~\cite{sharma2012novel}.

In the testing set, we use the chosen threshold of $0.00368$ to detect malicious traffic data. We illustrate the efficacy of our proposed approach by reporting the detection rate of the five different traffic types. The detection rate is the fraction of the detected traffic for an individual traffic type. Table~\ref{table:detect} summarizes the detection rate for the five different traffic types. In the case of normal traffic data, the observations less than the threshold are considered a positive detection. On the other hand, for the other four types, the observations that have a reconstruction error greater than the threshold are considered a positive detection. 

Table~\ref{table:detect} clearly shows that the detection rate is higher for a particular error type. This indicates that for a considered traffic type, most of its observations are detected properly. However, in this paper, we focus our attention on a binary classification problem and do not delve further into classifying the various types of attacks. The four anomalous traffic data -- DoS, R2L, U2R and probe are categorised in a single category and is termed as attack.


\begin{table}[htb]
\centering
\normalsize 
\begin{tabular}{l|c}
\textbf{Traffic Type} & \textbf{Detection Rate (\%)} \\
\hline 
Normal & 88.2 \\
DoS & 92.5 \\
R2L & 100.0 \\
U2R & 100.0 \\
Probe & 99.1\\
\hline 
\end{tabular}
\caption{Detection rate of the five different traffic types. This value is obtained in the testing set of the NSL-KDD dataset.}
\label{table:detect}
\end{table}

\begin{figure*}
\centering
\subfloat[SVM]{\includegraphics[height=1.4in]{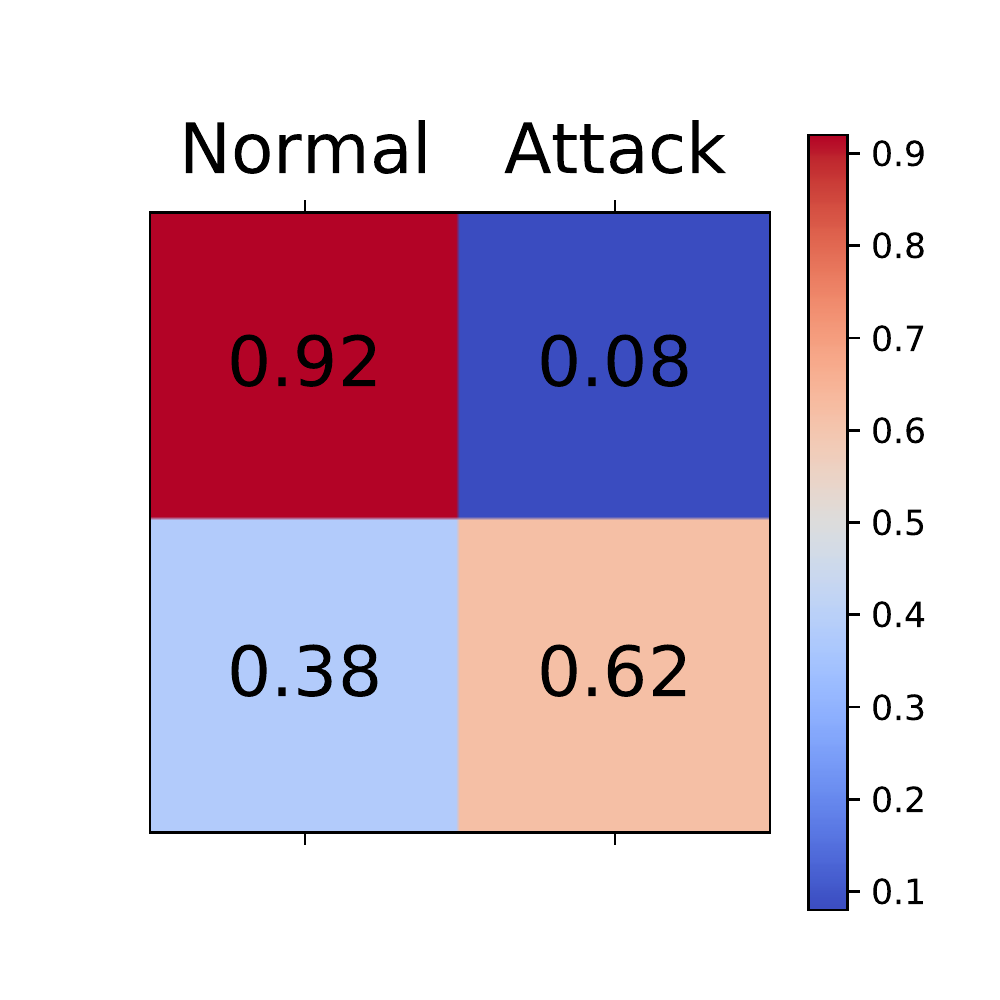}}
\subfloat[J48]{\includegraphics[height=1.4in]{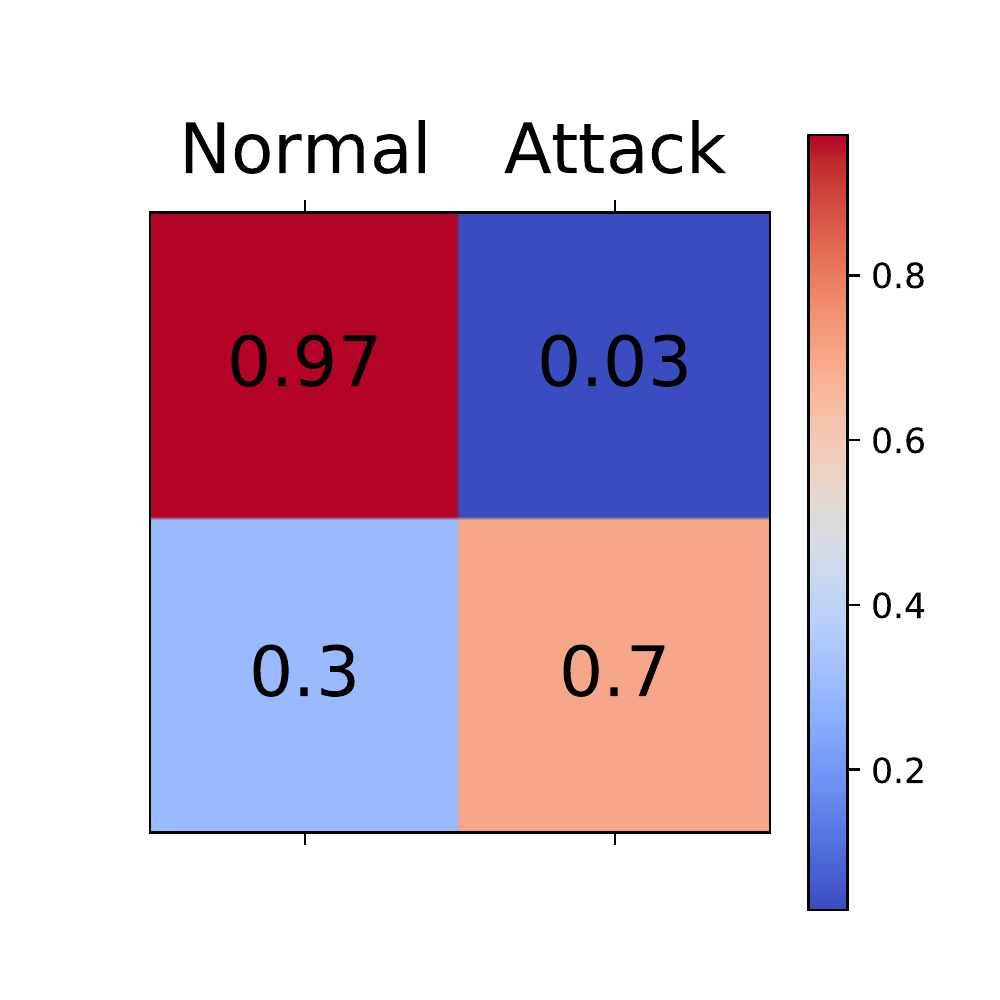}}
\subfloat[Naive Bayes]{\includegraphics[height=1.4in]{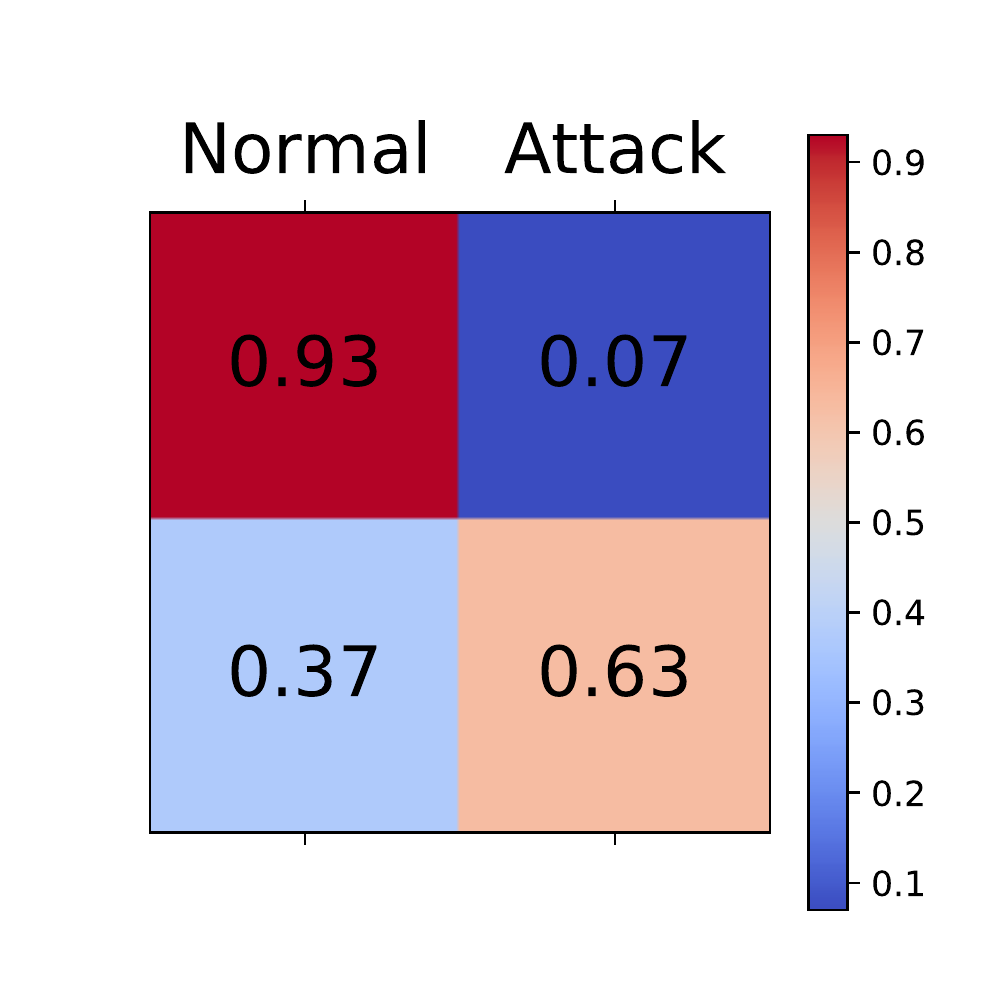}}
\subfloat[Random Forest]{\includegraphics[height=1.4in]{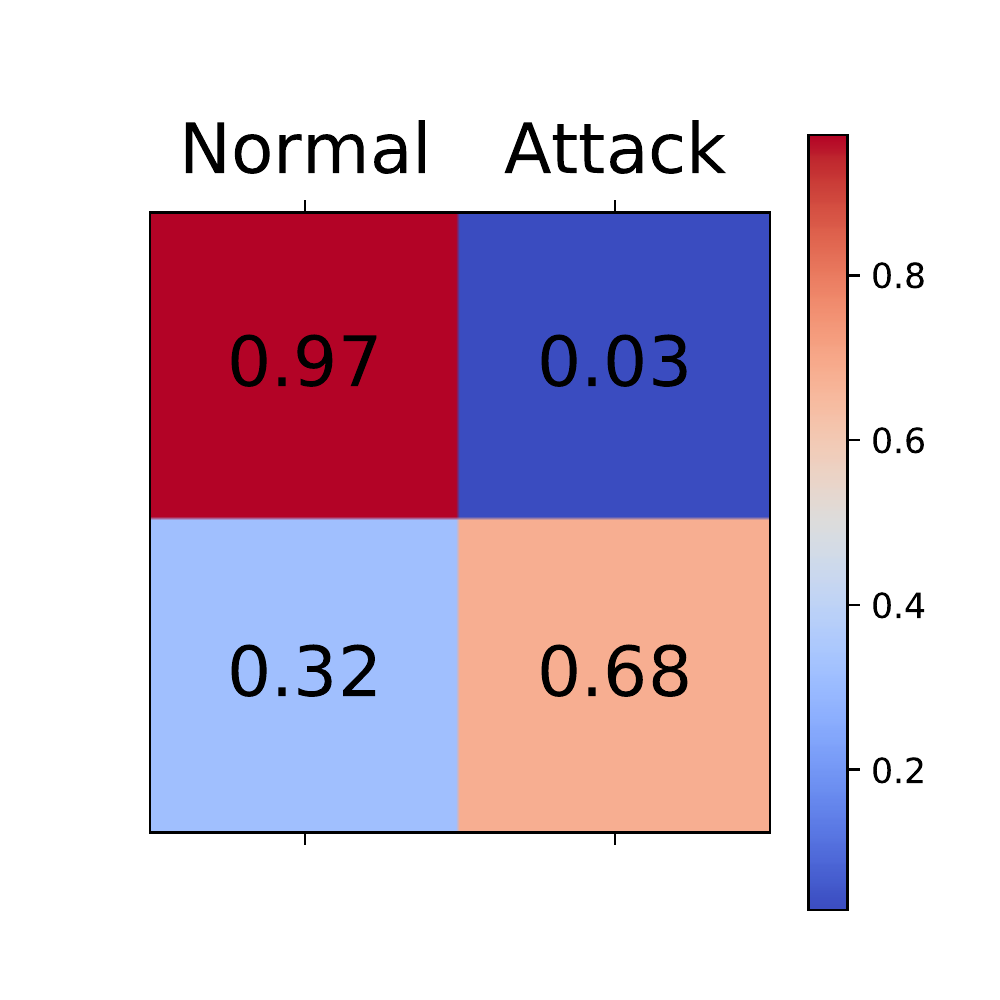}}
\subfloat[Proposed approach]{\includegraphics[height=1.4in]{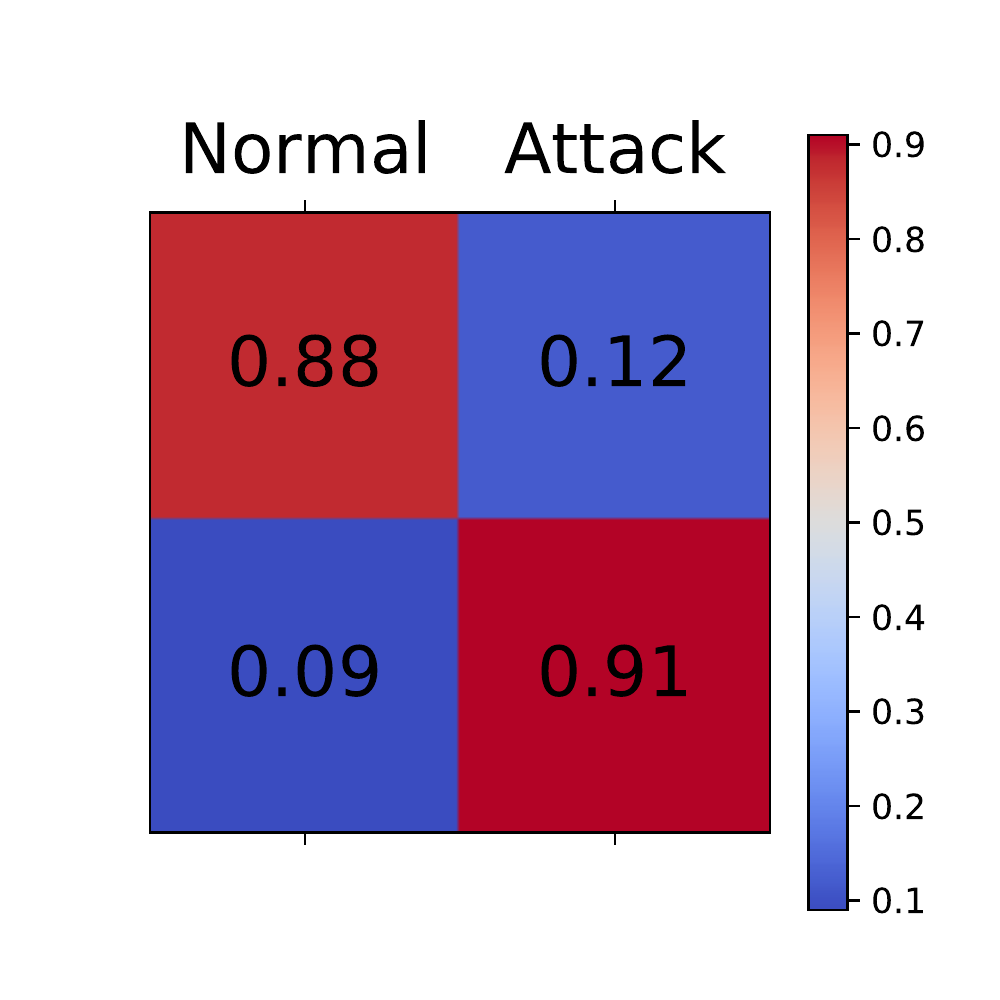}}
\caption{Confusion matrix obtained from our proposed approach, and other benchmarking methods. We observe that our proposed approach is better in correctly identifying the attacks, as compared to the other methods.}
\label{fig:confusion-matrix}
\end{figure*}

We also check the efficacy of our proposed method for the classification of normal data and attack. Figure~\ref{fig:confusion-matrix} illustrates the confusion matrix that we obtain in the testing stage using our proposed DL approach. We also show the confusion matrices obtained from the other traditional ML approaches. Each event in the testing set either belong to \emph{normal} or \emph{attack} events. We observe that most of the benchmarking methods have a poor detection rate for the attack traffic. Furthermore, our proposed approach has a high degree of accuracy of $0.91$ in correctly detecting the \emph{attack} events in the network. However, our accuracy is slightly lower at $0.88$ for \emph{normal} traffic events. This is reasonable for all practical applications, as it is important to correctly identify all the attacks in network traffic. 




\subsection{Receiver Operating Characteristic (ROC)}
\begin{figure}[htp]
  \begin{center}
    \includegraphics[width=2.5in]{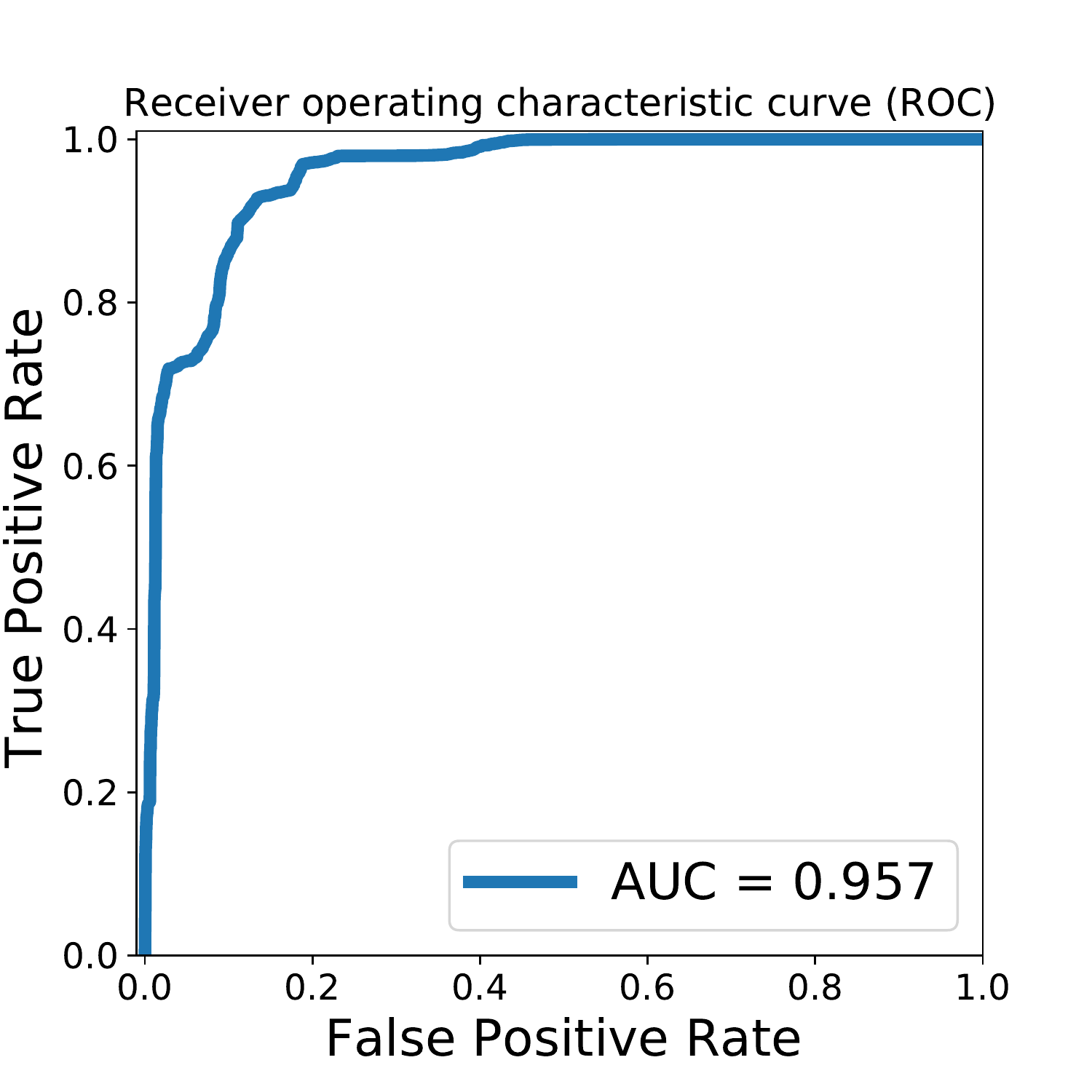}
  \end{center}
  \caption{Receiver Operating Curve (ROC) of our proposed approach across varying threshold values.}
  \label{fig:roc}
\end{figure}

The proposed method is dependent on the choice of the threshold for detecting malicious traffic data. We use the ROC curve in order to demonstrate the impact of the threshold on the performance of our proposed approach. The ROC curve is the plot between false positive rate and true positive rate, for varying thresholds. Figure~\ref{fig:roc} describes the ROC curve of our proposed method.

The area under the ROC curve, popularly known as area Under the Curve (AUC), which is a measure of the efficacy degree of the binary classifier. A ROC curve with a unity AUC indicates the perfect binary classifier. We compute the AUC value for Figure~\ref{fig:roc}, and obtain the value of $0.957$. This high value indicates the good performance of our proposed method.

\subsection{Benchmarking}
Finally, we benchmark our proposed method with other state-of-art classification models. We used precision, recall, f-Score, and detection rate, to evaluate our model performance. 
The mathematical representation of these metrics are calculated as follows:

\begin{equation}
	\mbox{Precision}=\frac{TP}{TP+FP}
    \label{eq1}
\end{equation}   

\begin{equation}
	\mbox{Recall}=\frac{TP}{TP+FN}
    \label{eq2}
\end{equation}   

\begin{equation}
	\mbox{F-score}=\frac{2\times\mbox{Precision}\times\mbox{Recall}}{\mbox{Precision}+\mbox{Recall}}
    \label{eq3}
\end{equation}   

\begin{equation}
	\mbox{Accuracy}=\frac{TP+TN}{TP+TN+FP+FN}
    \label{eq4}
\end{equation} 

where TP (True Positive) represents the number of instances correctly classified as an attack; TN (True Negative) represents the number of instances correctly classified as normal; FP (False Positive) represents the number of instances incorrectly classified as an attack; FN (False Negative) represents the number of instances incorrectly classified as normal.

\begin{table*}[htb]
\centering
\begin{tabular}{p{4cm}||l|c|c|c|c}
\textbf{Dataset} & \textbf{Approach} & \textbf{Precision (in \%)} & \textbf{Recall (in \%)} & \textbf{F-score (in \%)} & \textbf{Accuracy (in \%)} \\
\hline
\multirow{5}{*}{KDDTrain+ and KDDTest+} & SVM~ & 80 & 75 & 75 & 75.3 \\
 & J48~ & 85 & 81 & 81 & 81.5 \\
 & Naive Bayes~ & 80 & 76 & 75 & 76.1 \\
 & Random Forest~ & 85 & 80 & 80 & 80.4 \\
 & Latah \textit{et al.}~\cite{latah2018efficient}~ & 94 & 77 & 85& 84.29\\
 & Prasath \textit{et al.}~\cite{prasath2019meta}~ & 77& 74 & 75 & 82.99\\
 & Tang \textit{et al.}~\cite{tang2016deep}~ & 83 & 76 & 75& 74.67\\
 & \textbf{Proposed Model}  & \textbf{90.99} & \textbf{90.51} &  \textbf{90.75} &  \textbf{89.49} \\
\hline 
\end{tabular}
\caption{Benchmarking of our proposed method with other state-of-art network intrusion detection techniques. We report the precision, recall, F-score and accuracy for the different benchmarking algorithms. We report the results by training on KDDTrain+, and testing on KDDTest+ dataset.}
\label{table:benchmarking}
\end{table*}

For a comprehensive evaluation of the proposed model, we compared our model with some of the most well-used ML algorithms, also with some of the most well-known techniques that have employed the NSL-KDD dataset for their evaluation.
Firstly, we benchmark against classical ML algorithms such as SVM, J48~, Naive Bayes (NB) and Random Forest (RF). Secondly, we compare our proposed model with the existing state of the art  techniques,including~\cite{latah2018efficient}\cite{prasath2019meta}\cite{tang2016deep}. 
We analyse the precision, recall, F-score, and accuracy values for all methods. The experiments are conducted by training the model on the full dataset KDDTrain+ and testing it on the full dataset KDDTest+. The results are presented in Table~\ref{table:benchmarking}. We noticed that our approach has the best performance metrics compared to various classification techniques. The other techniques analysed including~\cite{latah2018efficient}\cite{prasath2019meta}\cite{tang2016deep} fail to have competitive F-score and accuracy values.

\section{Conclusion and Future Work}
\label{sec:conc}

The network data can often be compromised by malicious attacks by intruders. Therefore, researchers usually deploy machine learning based frameworks to detect anomalies during cybersecurity attacks. In this paper, we highlight the existing problems in related methods and propose solutions to address them. We propose a deep learning approach based on LSTM-autoencoder that can model the normal traffic data efficiently. Our experiments indicate that the proposed model can accurately detect the anomalies presented in network traffic data. In our future work, we plan to apply our current model on various benchmark datasets, such as CICIDS 2017, Kyoto, UNSW-NB15, and ISC2012. We also plan to extend the binary classification problem into a multi-class classification problem. Such that the type of network attacks are also identified and categorized accurately.

\balance

\bibliographystyle{IEEEtran}

\end{document}